# Unexpectedly Spontaneous Water Dissociation on Graphene Oxide Supported by Copper Substrate

Zhijing Huang[*, a], Zihan Yan[*, a], Guangdong Zhu[a], Xing Chen[a], Shuming Zeng[a], Xiuyun Zhang[a], Liang Zhao[†, a] and Yusong Tu[†, a]

[a]*College of Physical Science and Technology, Yangzhou University, Jiangsu, 225009, China.*
[*]Authors contribute equally to the paper.
[†]Corresponding author. Email: zhaoliang@yzu.edu.cn; ystu@yzu.edu.cn

**Abstract**
Water dissociation is of fundamental importance in scientific fields and has drawn considerable interest in diverse technological applications. However, the high activation barrier of breaking the O-H bond within the water molecule has been identified as the bottleneck, even for the water adsorbed on the graphene oxide (GO). Herein, using the density functional theory calculations, we demonstrate that the water molecule can be spontaneously dissociated on GO supported by the (111) surface of the copper substrate (Copper-GO). This process involves a proton transferring from water to the interfacial oxygen group, and a hydroxide covalently bonding to GO. Compared to that on GO, the water dissociation barrier on Copper-GO is significantly decreased to be less than or comparable to thermal fluctuations. This is ascribed to the orbital-hybridizing interaction between copper substrate and GO, which enhances the reaction activity of interfacial oxygen groups along the basal plane of GO for water dissociation. Our work provides a novel strategy to access water dissociation via the substrate-enhanced reaction activity of interfacial oxygen groups on GO and indicates that the substrate can serve as an essential key to tuning the catalytic performance of various two-dimensional material devices.

## 1. Introduction

Water dissociation is of fundamental importance in various physical, chemical and biological processes, such as interfacial wetting[1, 2], metal corrosion[3, 4]and carbon dioxide equilibria [5, 6], and has drawn considerable interest in renewable energy storage[7, 8], fuel cells manufacture[9, 10]and the preparation of liquid hydrocarbons[11, 12]. The key principle of water dissociation is to break the O-H bond by increasing input energy or reducing the reaction barrier. For instance, the direct thermolysis route[13, 14] works in an extremely high-temperature environment, whereas the electro- or photo-chemical[15-18] approaches commonly introduce catalysts to reduce the water dissociation barrier. For practical applications, it is essential to develop simpler and more reliable strategies with low energy consumption and high reaction efficiency. Despite many efforts devoted to the routes of water dissociation in diverse systems, including metal or metal oxides[19, 20], single or multilayered carbon nanotubes[21, 22] and doped or defective graphene[23, 24], it is still a great challenge to achieve water dissociation at ambient conditions.

Graphene oxide (GO) is a carbon-based material covalently functionalized by oxygen-containing groups dominated by epoxy and hydroxyl on its basal plane[25-29] and it is usually fabricated on metal substrates in experiments by the chemical vapor deposition method [30, 31]. The presence of metal substrates can alter the interfacial adsorption[32, 33], electronic[34, 35] and thermal transport properties[36] of GO. Of all the metal substrate, copper substrate is the most used due to the low solubility of carbon atoms in copper compared to other metal substrates[37, 38], and its (111) surface is found to assist the fast growth of stable and large-area GO because of the well-matched lattice between the Cu(111) surface and GO[39, 40] [41]. This advantage makes the copper substrate-supported GO (Copper-GO) a highly promising material for technological applications such as coatings against aqueous corrosion [42], batteries [43] and humidity sensors [44]. In particular, it has been demonstrated that the copper substrate remarkably enhances the dynamic migration of oxygen groups on GO [41], which represents the enhanced activity of oxygen-containing groups on GO. It is expected this enhanced activity of oxygen groups should catalyze the oxygen-related reactions occurring at the interface of GO, such as water dissociation.

In this work, we have performed density functional theory (DFT) calculations to explore the water dissociation process on Copper-GO. Unexpectedly, we find that the water molecule can be spontaneously dissociated on Copper-GO, along with a proton transferring from water to the interfacial oxygen group, and a hydroxide covalently bonding to GO. To the best of our knowledge, this is the first report of the copper substrate-enhanced reaction activity of interfacial oxygen groups on GO to access the water dissociation under ambient conditions.

## 2. Model and Methods

The simulation system consists of a water molecule adsorbed on a GO sheet with or without the supported copper substrate. Their periodic boxes have *xyz* dimensions of 12.78 ×12.78×25.0 Å$^3$ in a 5×5 supercell. To construct the periodic Copper-GO system, the lattice of GO is stretched by 3.7% to match the (111) surface of the copper substrate. The lowest layer in the four-layers copper substrate is fixed in the DFT calculations. A vacuum layer of about 15 Å is used to avoid interactions between the periodic cells in the *z*-axis direction. All the DFT calculations were carried out with the Vienna ab initio simulation package (VASP)[45]. Given that the generalized gradient approximation (GGA) usually overestimates the distance between the graphene and metal substrates even with the vdW correction[32, 46, 47], we use the local density approximation (LDA) to describe the exchange-correlation energy, which has been proved to be more reasonable for the system of graphene supported by metal substrates[32, 47, 48]. The Perdew–Burke–Ernzerhof (PBE) functional with optB86b-vdW correction is also employed to check the barrier of the water dissociation on GO and Copper-GO, and the results are consistent with those using LDA (see PS. 1 of Supporting Information (SI)). The projector augmented wave (PAW) is used with the plane wave cutoff of 500 eV. The convergence of energy and force are $10^{-5}$ eV and 0.01 eV/Å, respectively. The Monkhorst-Pack scheme[49] of 1×1×1 *k*-point mesh in the Brillouin zone is used. After optimizing the reactant and product, we search the transition state (TS) via the dimer method[50], and only one imaginary frequency affirms the true TS structure.

## 3. Results and Discussion

### 3.1 Water Dissociation on GO and Copper-GO

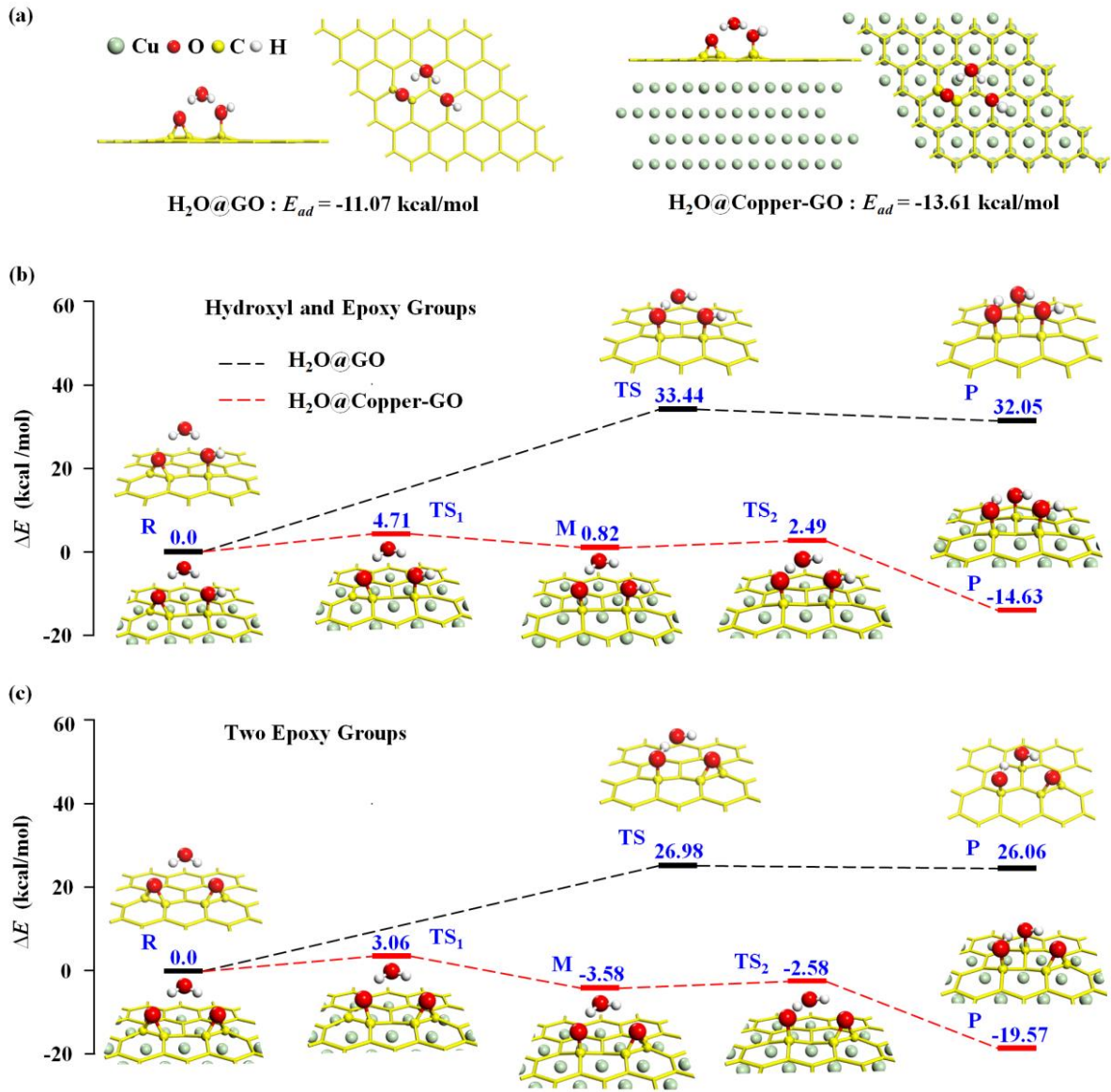

**Figure 1.** (a) Top and side snapshots of a water molecule adsorbed on GO and Copper-GO with hydroxyl and epoxy groups. (b) Reaction pathways of water dissociation on GO (black lines) and Copper-GO (red lines) for hydroxyl and epoxy groups and (c) two epoxy groups. Notations: R (reactant), M (intermediate), TS (transition state) and P (product). The energy values of R for GO and Copper-GO are shifted to 0 for easy comparison.

Figure 1(a) shows the representative configurations of dominating oxygen groups with water adsorbed on GO and Copper-GO. We choose the correlated distribution of oxygen groups, i.e., a pair of hydroxyl and epoxy groups, two epoxy groups and two hydroxyl groups, since the oxygen groups on the basal plane of GO have shown to be distributed in correlation, which is consistent with the observations that the coexistence of both unoxidized and oxidized regions[27]. The adsorption energies of a water molecule on GO or Copper-GO, $E_{ad}$, is defined as $E_{ad} = E_{H_2O@GO/Copper-GO} - E_{GO/Copper-GO} - E_{H_2O}$, where $E_{H_2O@GO/Copper-GO}$, $E_{GO/Copper-GO}$ and $E_{H_2O}$ are the energies of the corresponding optimized structures, respectively. We found that the $E_{ad}$ on GO and Copper-GO with hydroxyl and epoxy groups are -11.07 kcal/mol and -13.61 kcal/mol, respectively; the $E_{ad}$ on GO and Copper-GO with two epoxy groups are -9.45 kcal/mol and -12.45 kcal/mol (see PS. 2 in SI for the adsorption configuration and energy), respectively; the $E_{ad}$ on GO and Copper-GO with two hydroxyl groups are -9.22 kcal/mol and -13.37 kcal/mol, respectively (see PS. 3 in SI). These lower values of $E_{ad}$ on Copper-GO indicate that the water molecule is more favorably adsorbed on Copper-GO.

The presence of copper substrate significantly decreases the water dissociation barrier on GO with hydroxyl and epoxy groups. Fig. 1(b) shows the reaction pathways of the water dissociation process on GO and Copper-GO. We can see that the water dissociation is accessed by two successive steps: a proton transfers from water to the interfacial oxygen group, and a hydroxide covalently bonds to GO. On GO, there is a much higher reaction barrier of 33.44 kcal/mol, preventing water dissociation under ambient conditions[51]. In contrast, surprisingly, on Copper-GO, the reaction barrier is significantly reduced, down to 1.67 kcal/mol. An intermediate with a dangling C-O bond is formed after a C-O bond in epoxy breaks with a reaction barrier of 4.71 kcal/mol. These energy barriers less than or comparable to thermal fluctuations indicate that water can be spontaneously dissociated on Copper-GO at ambient conditions. It should be emphasized that the existence of an intermediate with a dangling C-O bond indicates that the enhanced reaction activity of interfacial oxygen groups by the copper substrate provides possible active sites in water dissociation on Copper-GO.

Figure 1(c) further demonstrates the sharply decreasing of water dissociation barrier on Copper-GO with two epoxy groups. Similar to the case on Copper-GO with hydroxyl and epoxy groups, the reaction barrier is also sharply reduced, from 26.98 kcal/mol on GO to 1.0 kcal/mol on Copper-GO. The intermediate with a dangling C-O bond also occurs and the C-O bond breaking reaction presents a much lower barrier of 3.06 kcal/mol. For Copper-GO with two hydroxyl groups, the water dissociation barrier is found to be 9.45 kcal/mol (see the reaction pathways in PS. 4 in SI). These energy barriers on Copper-GO less than or comparable to thermal fluctuations confirm that water can be spontaneously dissociated on Copper-GO at ambient conditions.

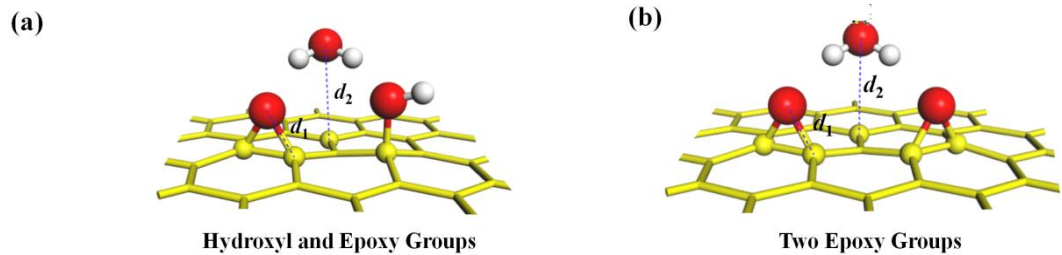

(a) Hydroxyl and Epoxy Groups

(b) Two Epoxy Groups

(c) The distances between atoms for states along the water dissociation paths on GO and Copper-GO, in the unit of Å.

|  | states | Hydroxyl and Epoxy Groups | | Two Epoxy Groups | |
| --- | --- | --- | --- | --- | --- |
|  |  | $d_1$ | $d_2$ | $d_1$ | $d_2$ |
| GO | R | 1.441 | 2.956 | 1.444 | 2.940 |
|  | TS | 1.431 | 2.007 | 1.433 | 1.920 |
|  | P | 1.453 | 1.483 | 1.455 | 1.481 |
| Copper-GO | R | 1.469 | 2.788 | 1.469 | 2.727 |
|  | $TS_1$ | 1.396 | 2.709 | 1.404 | 2.658 |
|  | M | 1.343 | 2.590 | 1.342 | 2.515 |
|  | $TS_2$ | 1.362 | 2.049 | 1.368 | 2.049 |
|  | P | 1.424 | 1.455 | 1.422 | 1.459 |

**Figure 2.** Schematic illustrations of C-O distances ($d_1$ and $d_2$) for GO and Copper-GO with (a) hydroxyl and epoxy groups and (b) two epoxy groups. (c) Values of $d_1$ and $d_2$ in various states along the water dissociation paths on GO and Copper-GO. Notations are the same as those in Fig. 1.

Figure 2 shows the variation of relevant bond distances along the water dissociation paths on GO and Copper-GO that indicate the enhanced reaction activity of interfacial oxygen groups on Copper-GO. On GO and Copper-GO with hydroxyl and epoxy groups (Fig. 2(c)), the length of the C-O bond in epoxy (denoted by $d_1$) in R on Copper-GO (1.469 Å) is longer than that on GO (1.441 Å), indicating that the C-O bond is elongated and becomes more active on Copper-GO. In its $TS_1$, M and $TS_2$, an intermediate with the dangling C-O bond appear and $d_1$ is all shortened on Copper-GO than on GO. Water is getting closer to GO in the dissociation process, and the water-GO distance $d_2$ decreases, finally to the minimum of 1.483 Å (on GO) and 1.455 Å (on Copper-GO) in P after the water is dissociated. Making the comparisons of water dissociation on GO and Copper-GO with two epoxy groups or two hydroxyl groups, we also see similar variations of $d_1$ and $d_2$. The distances between water and two oxygen groups are listed in PS. 5 of SI, indicating the enhanced water-GO interaction on Copper-GO.

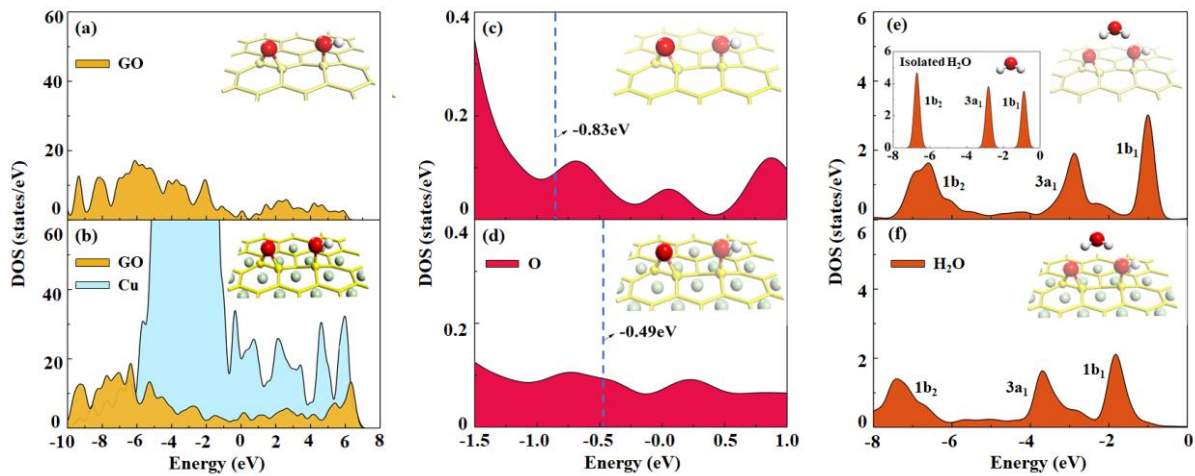

**Figure 3.** Density of states (DOS) of GO (top) and Copper-GO (bottom) with hydroxyl and epoxy groups. (a, b) Summed DOS of *s*, *p* orbitals of carbon and oxygen, and *s* orbital of hydrogen of GO (yellow), as well as *s*, *p* and *d* orbitals of the copper substrate (cyan), on GO and Copper-GO; (c, d) DOS of the *p* orbital of oxygen atom in epoxy (O, red) on GO and Copper-GO. The blue dash lines represent the *p*-band centers of the oxygen atom in epoxy at relevant energy levels; (e, f) DOS of the bonding orbitals of water ($H_2O$, orange) adsorbed on GO and Copper-GO. Inset in (e) presents three bonding orbitals of isolated water, denoted by $1b_2$, $3a_1$ and $1b_1$, respectively.

The analysis of the density of states (DOS) demonstrates that the orbital-hybridizing interaction between copper substrate and GO enhances the reaction activity of interfacial oxygen groups along the basal plane of GO for water dissociation. Fig. 3(a) and 3(b) show the DOS of free-standing GO and Copper-GO with hydroxyl and epoxy groups. The DOS of GO overlaps largely with these of copper substrate on Copper-GO, which indicates the occurrence of orbital-hybridizing interaction between copper substrate and GO. Especially as shown in Fig. 3(c) and 3(d), around the Fermi level, we can see that while the *p*-band center of the oxygen atom in epoxy on free-standing GO is at -0.83eV, the center on Copper-GO shifts rightward to -0.49 eV. Near the Fermi level, the number of electronic states of the oxygen atom in epoxy (epoxy oxygen) on Copper-GO increases significantly compared to free-standing GO. This indicates this orbital-hybridizing interaction between copper substrate and GO enhances the reaction activity of epoxy oxygen on GO. Fig. 3(e) and 3(f) show the DOS of the bonding orbitals ($1b_2$, $3a_1$ and $1b_1$) of water adsorbed on GO and Copper-GO. We can see that the peaks of these three orbitals of water adsorbed on GO are decreased and the orbitals are broadened compared to the orbitals of isolated water (see the Inset of Fig. 3(e)), and the peaks of the orbitals of water adsorbed on Copper-GO are further lowered and the orbitals become wider. This indicates that the O-H bond within the water molecule is weakened and easy to dissociate. Moreover, on GO and Copper-GO with two epoxy groups, the *p*-band center of the epoxy oxygen also shifts toward the Fermi level from -0.68eV to -0.47 eV, respectively, and on Copper-GO the DOS around the Fermi level is found to significantly increase and the peaks of the orbitals of water adsorbed also are further lowered and the orbitals become wider, compared with the DOS on GO (see the details in PS. 6 in SI). For two hydroxyl groups, the *p*-band center of hydroxyl oxygen shifts from -0.91 eV to -0.75 eV, and DOS show the similar changes to those for two epoxy groups or a pair of hydroxyl and epoxy groups (see the details in PS.7 in SI). All results above demonstrate that the orbital-hybridizing interaction between copper substrate and GO enhances the reaction activity of interfacial oxygen groups along the basal plane of GO and thus leads to a significant decrease in the barriers of water dissociation.

**3.2 Water Dissociation on GO Supported by Other Metal Substrates**

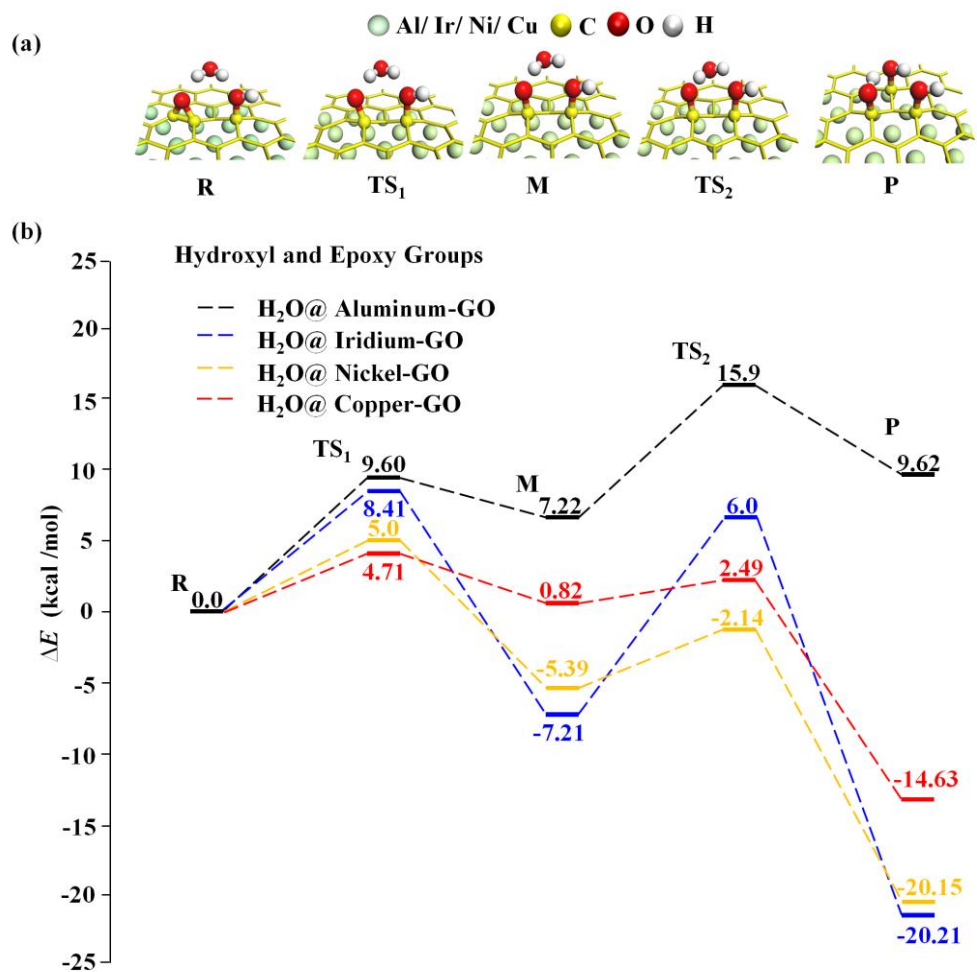

**Figure 4.** (a) Schematic configurations of various states in the pathways of water dissociation for a pair of hydroxyl and epoxy groups on GO supported by the (111) face of metal substrates including aluminum (Al), iridium (Ir), nickel (Ni) and copper (Cu). (b) Comparison of the water dissociation pathways on various metal substrates-supported GO. The result for Copper-GO is the same as that in Fig. 1(b). The energy values of R are shifted to 0 for easy comparison.

To investigate the effect of metal substrates on water dissociation, we compare the reaction pathways of water dissociation for a pair of hydroxyl and epoxy groups on GO, supported by the (111) face of aluminum, iridium, nickel and copper substrates. As shown in Fig. 4(a), all of these systems show similar water dissociation processes involving two successive steps, i.e., breaking of the C-O bond in epoxy and proton transfer from water to the interfacial oxygen group. Fig. 4(b) further compares the reaction pathways of water dissociation process. For the C-O bond breaking reaction, we can see that Copper-GO shows the lowest barrier of 4.71 kcal/mol while Aluminum-GO has the highest barrier of 9.60 kcal/mol. This suggests that epoxy on Copper-GO can be more easily converted to a dangling C-O bond. For the proton transfer, the highest barrier is 13.21 kcal/mol for Iridium-GO and Copper-GO still shows the lowest barrier of 1.67 kcal/mol. These findings demonstrate that water dissociation occurs more easily on Copper-GO than on other metal substrates-supported GO, thus confirming that the presence of copper substrate significantly reduces the reaction barrier of water dissociation on GO.

**3.3 Water Dissociation for Oxygen Groups with Varied Separations**

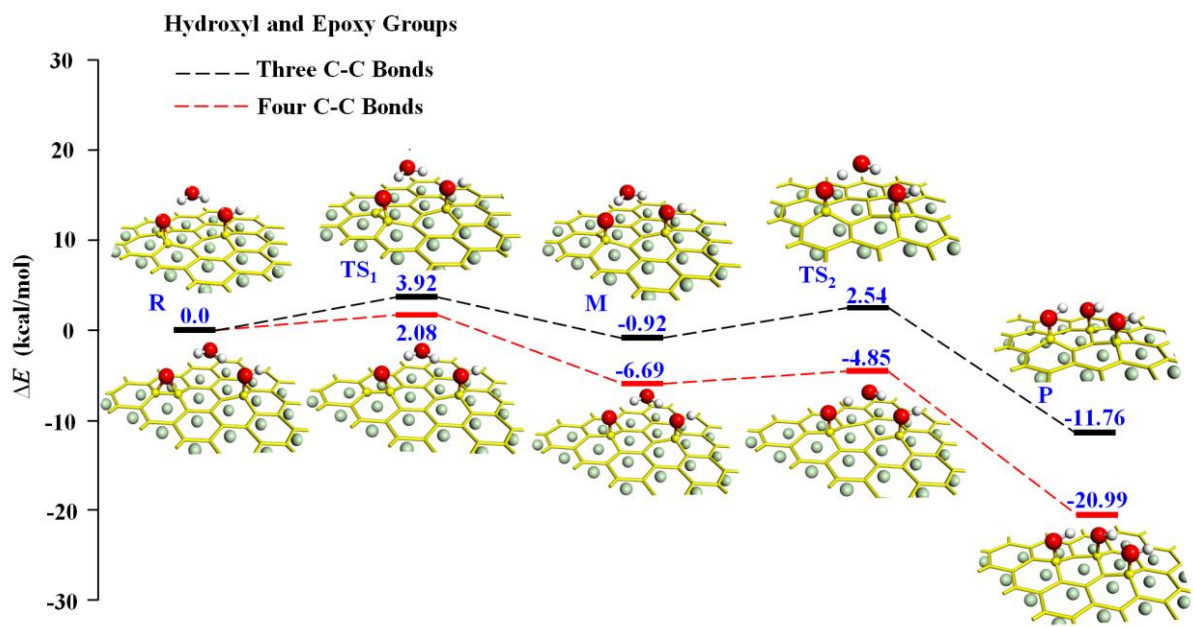

**Figure 5.** Reaction pathways of water dissociation on Copper-GO with the varied inter-distance between hydroxyl and epoxy groups: (a) three and (b) four C-C bonds separation. Notations are the same as those in Fig. 1.

The appropriate separation between oxygen groups benefits the water dissociation reactions on Copper-GO. In Fig. 5, as the hydroxyl and epoxy groups are separated by three C-C bonds separation on Copper-GO, the reaction barriers of C-O bond breaking and proton transfer are found to be 3.92 kcal/mol and 3.46 kcal/mol. Compared to two C-C bonds separation (see Fig. 1(b)), these reaction barriers are lower than 4.71 kcal/mol and larger than 1.67 kcal/mol. As the inter-distance increases to four C-C bonds separation, the reaction barriers are reduced to 2.08 kcal/mol and 1.84 kcal/mol, respectively. When the hydroxyl and epoxy groups are separated into five or more C-C bonds, the water molecule cannot be accommodated between these two oxygen groups by hydrogen bonds. The water dissociation barrier shows similar behaviors on Copper-GO with two epoxy groups or two hydroxyl groups (see PS. 8 and PS. 9 in SI), and three or two C-C bonds separation corresponds to the most appropriate separation

**4. Conclusions and discussion**

In summary, we have demonstrated that the water molecule can be spontaneously dissociated on Copper-GO, i.e., a proton transfers from water to the interfacial oxygen group, and a hydroxide covalently bonds to GO. It should be noted that water dissociation on free-standing GO has a much higher reaction barrier, while the barrier on Copper-GO is reduced to be less than or comparable to thermal fluctuations. The significant decrease of the water dissociation barrier is attributed to the orbital-hybridizing interaction between copper substrate and GO, which enhances the reaction activity of interfacial oxygen groups along the basal plane of GO for water dissociation. Compared to aluminum, iridium and nickel substrates-supported GO, the copper substrate-supported GO shows the lowest water dissociation barrier. We also found that the appropriate separation between oxygen groups can offer favorable H-bonding configurations with water and benefits the water dissociation on Copper-GO.

It should be noted that the conclusion may seemingly contradict with the study by David and Kumar[52], who show that GO can split water without a copper substrate. We have compared the model, the reaction conditions and the computational methods in our work with those used in David and Kumar's work. For the model, there are 265 water molecules in their system while there is only 1 water molecule in our system. The ratio of C/O on GO in their work is 4:1, but it is only 25:1 in our work, *i.e.*, the oxidation degree of GO in their system is higher than that of GO in our system. Previous studies have shown that the number of water molecules[53] and the oxidation degree[54] can affect the interaction between water molecules and GO interface. For the reaction conditions, the temperature in their work is 300 K, but it is 0 K in our work. The temperature will also influence the behavior of water molecules on the GO interface[55]. For the computational methods, the *ab initio* molecular dynamics (AIMD) simulations were performed in their work, whereas we performed the time-independent first-principles calculations in our work. The former method gives the time-dependent evolution of water-GO interface, whereas ours focuses on the search of water dissociation pathways. Therefore, we consider that the water dissociation depends on many factors, such as the adsorption of more water molecules[53], change of oxidation degree of GO[52, 54] or reaction temperature, and the two investigations are based on different environment conditions.

Our work provides a possible application of Copper-GO system in the coatings against aqueous corrosion. GO is a promising coating material against aqueous corrosion as the presence of GO significantly reduces the corrosion rate of copper metal[42]. Indeed, the bare copper metal can be easily corroded in the aqueous environment since the barrier of water dissociation on the copper surface is only 3.92 kcal/mol under ambient conditions[56]. However, when a layer of GO is deposited on the surface of copper metal, the direct contact between water molecules and copper metal is effectively avoided, preventing the copper metal from being corroded effectively[57]. Furthermore, our calculations have shown that the barrier for water dissociation on Copper-GO can be further lowered to 1.67 kcal/mol. The decrease of water dissociation barrier implies that the water molecule is much easier to be dissociated on Copper-GO than on the surface of copper metal.

In general, traditional water dissociation routes, such as doping, light irradiation and electrolysis, usually rely on precious metals[58], ultraviolet irradiation[51] and high overpotential[59], which are difficult to realize under ambient conditions. Considering that pristine graphene has been fabricated with a large coverage area on the copper substrate in experiments, the molecular understanding of the substrate-enhanced reaction activity of interfacial oxygen groups on GO presented here offers crucial insights for GO-relevant desirable designs and potential applications.

**Author contributions**
Yusong Tu conceived, designed and guided the project. Zhijing Huang and Zihan Yan

performed the simulations. Zhijing Huang, Guangdong Zhu, Xing Chen, Shuming Zeng and Xiuyun Zhang analyzed the data. Zhijing Huang and Liang Zhao wrote the paper. All authors approved the final version of the manuscript.

**Conflicts of Interest**
The authors declare that they have no known competing financial interests or personal relationships that could have appeared to influence the work reported in this paper.

**Acknowledgments**
We are thankful for the helpful suggestions given by Zonglin Gu, Zhaoju Gao, Hao Yang and Chengao Ji. This work was funded by the National Natural Science Foundation of China (Nos. 12075201, 11675138), the Natural Science Foundation of Jiangsu Province (No. BK20201428), the Special Program for Applied Research on Supercomputation of the NSFC-Guangdong Joint Fund (the second phase), Students' Innovation and Entrepreneurship Training Program of Yangzhou University (No. X20210261), Jiangsu Students' Innovation and Entrepreneurship Training Program (No. 202111117008Z) and Hefei Advanced Computing Center.